\documentclass[namedreferences]{solarphysics}
\usepackage{graphics,graphicx,multirow,cancel,url}
\usepackage{lscape}
\usepackage{ragged2e}
\usepackage[hyperref,optionalrh]{spr-sola-addons} 
\usepackage{color}           
\usepackage{breakurl}        
\usepackage{units} 
\usepackage{textcomp}
\usepackage{siunitx} 

\newcommand{\aap}{    {\it Astron. Astrophys.}}

\newcommand{\apj}{    {\it Astrophys. J.}}

\newcommand{\solphys}{{\it Solar Phys.}}
 
\newcommand{\ssr}{    {\it Space Sci. Rev.}}

\usepackage{epsfig}          
\usepackage{graphicx,color}        
\usepackage{amssymb} 
\usepackage{color}           
\usepackage{setspace}
\usepackage{rotating}
\usepackage{float}
\usepackage{textcomp}
\usepackage{natbib}
\begin{document}

\begin{article}

\begin{opening}

\title{Association of calcium network bright points with underneath photospheric magnetic patches}
\author{Nancy~\surname{Narang}$^1$\sep
	Dipankar~\surname{Banerjee}$^{1,2}$\sep
	 Kalugodu~\surname{Chandrashekhar}$^1$\sep
 Vaibhav~\surname{Pant}$^{3}$\sep}

\runningauthor{N.~Narang et al.}
\runningtitle{Association of network bright points with magnetic patches}
    \institute{$^{1}$ Indian Institute of Astrophysics, Koramangala 2nd Block, Bangalore, India
                     email: \url{nancy@iiap.res.in}}
                 
                  \institute{$^{2}$ Center of Excellence in Space Sciences, IISER, Kolkata, India
                 email: \url{dipu@iiap.res.in} }
             
              \institute{$^{3}$ Centre for Mathematical and Plasma Astrophysics, KU Leuven, Belgium}


\date{Accepted: XXXX , Received: \today}

\begin{abstract}

Recent dedicated  \textit{Hinode} polar region campaigns revealed the presence of concentrated kilogauss patches of magnetic field in the polar regions of the Sun which are also shown to be correlated with facular bright points at the photospheric level. In this work, we demonstrate  that this spatial intermittency of the magnetic field  persists even up to the chromospheric heights. The small-scale bright elements visible in the bright network lanes of solar network structure as seen in the Ca~{\sc ii~H} images are termed as network bright points. We use special \textit{Hinode} campaigns devoted to observe polar regions of the Sun to study the polar network bright points  during the phase of last extended solar minimum. We use Ca~{\sc ii~H} images of chromosphere observed by \textit{Solar Optical Telescope} (SOT). For magnetic field information, level 2 data of spectro-polarimeter (SP) is used. We observe a considerable association between the polar network bright points and magnetic field concentrations. The intensity of such bright points is found to be correlated well with the photospheric magnetic field strength underneath with a linear relation existing between them.

\end{abstract}

\keywords{Sun: polar regions, chromosphere, photosphere, magnetic fields, network structure, super-granulation}

\end{opening}


\section{Introduction}\label{sec:intro}

The chromospheric network is a web-like pattern most easily seen as emission feature in the red line of hydrogen ($\mathrm{H}\alpha$) and the violet lines of calcium (Ca~{\sc ii~K} and Ca~{\sc ii~H}) images of Sun. This pattern or network is believed to be coincident with the boundaries of large-scale convective cells, known as the supergranules, each about 20-60 Mm in diameter \citep{Rieutord10}. \citet{Simon64} found a one-to-one correlation in the position of supergranules and bright network seen on Ca~{\sc ii~K} line spectroheliograms (also see \citealp{Parker78, Sheeley11}). The supergranular convective flow across photosphere, pushes the magnetic elements towards the supergranules' boundaries, so a network of magnetic field is formed. These elements are also buffeted by exploding granules and sometimes get annihilated or grow in size. These magnetic field clumps with several hundreds of gauss at the photospheric level leads to enhanced heating in the solar atmosphere, with a consequent increased brightening at certain wavelengths in chromospheric and coronal heights.

\begin{figure}[htbp]
	
	\centering 
	
	\includegraphics[width=0.9\textwidth]{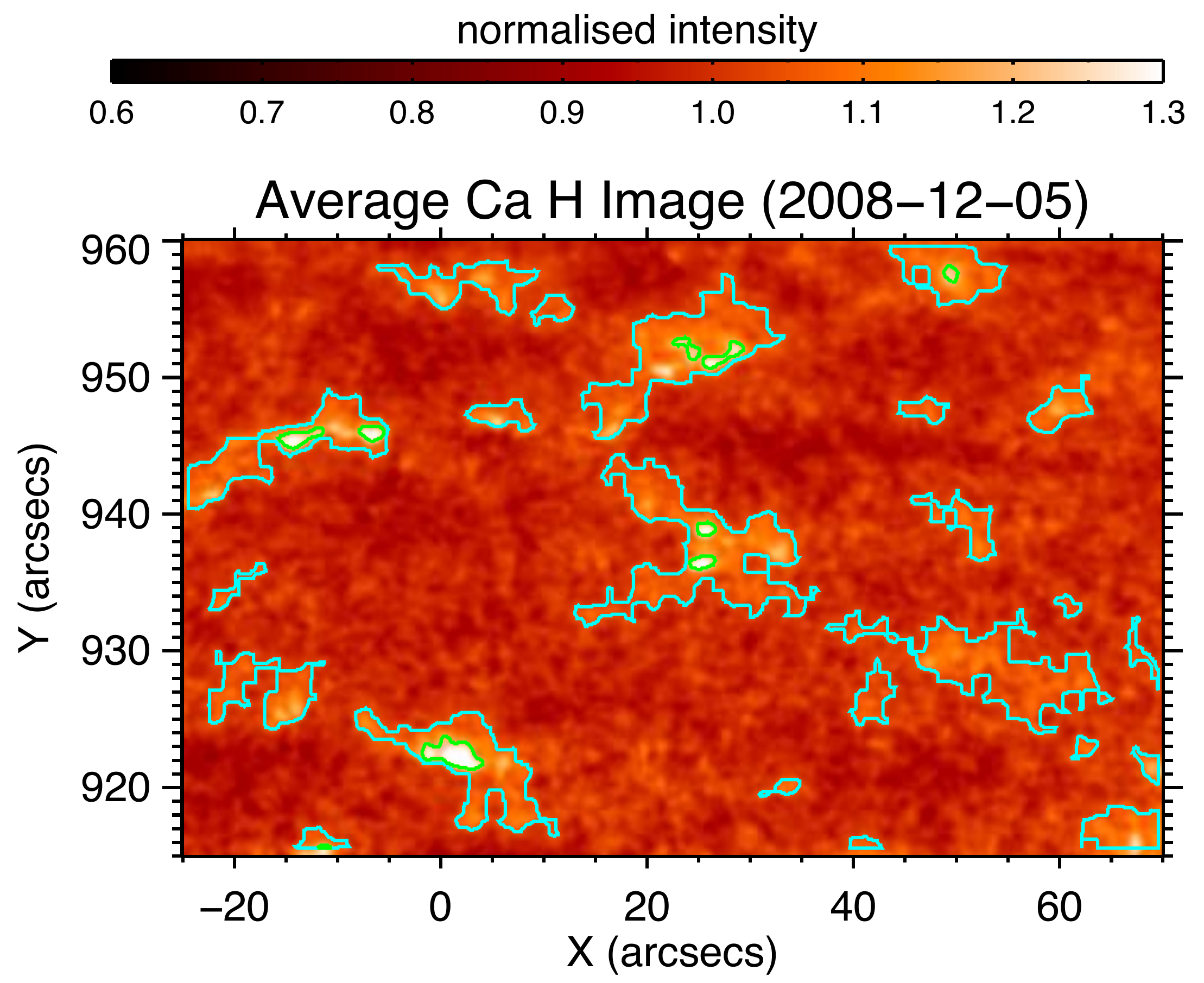}
	
	\caption{Representative example of observations from SOT which shows internetwork, network and network bright points. The \textit{blue contours} mark the locations of the bright network where the surrounding darker areas are called internetwork. The brighter locations present within the bright network are termed as network bright points which are marked by \textit{green contours}.}
	
	\label{fig1}
	
\end{figure}

\citet{Frazier70} has shown that the downflows at the vertices, where several supergranular cells meet are much more prominent for the concentration of magnetic flux than the rest of cell boundaries. At these points the magnetic field appears comparatively enhanced, resulting in coinciding network bright points recognisable in Ca~{\sc ii} lines. These bright points are brighter than rest of the network structure and proposed to be associated with strong ($\sim$1000 gauss) magnetic fields, probably in the form of magnetic flux tubes emerging from below the photosphere. Figure~\ref{fig1} shows an example of Ca~{\sc ii~H} observation obtained from \textit{Solar Optical Telescope} (SOT) \citep{Tsuneta08a} aboard \textit{Hinode} \citep{Kosugi07} which clearly marks the network bright points, and network and internetwork as distinguished features of the solar chromosphere. The nature of the network and its heating have been modeled by many authors (\textit{e.g.} \citealp{Gabriel76, Schrijver01, Wedemeyer09}). For instance, in the model by \citet{Gabriel76}, heating occurs along the field lines by magnetohydrodynamic (MHD) waves. Some oscillations may arise through the buffeting of narrow flux tubes at chromospheric altitudes by solar granulation, which leads to the propagation of MHD waves, causing the heating of chromospheric network and upper atmosphere. The heating is greatest where the field lines are most concentrated, at the common boundaries of many super-granules, in particular, network bright points. A different scenario involving magnetic dipole evolutions have also been proposed by many authors (\textit{e.g.} \citealp{Harvey73, Harvey85, Webb93}) for explaining the observed enhanced brightening in the different layers of the solar atmosphere. In this process, the change in magnetic flux due to cancellation, emergence, fragmentation, and coalescence occurring at the locations of supergranular network junctions leads to the excess chromospheric and coronal brightenings \citep{Egamberdiev83, Habbal90}.

%


Network bright points observed in polar regions of Sun are referred as polar network bright points. They populate higher heliographic latitudes above $60\,^{\circ}$ or $70\,^{\circ}$. Using high-resolution spectro-polarimetric observations with SOT, it is found that the polar region of Sun has isolated patches of the concentrated magnetic field with strengths exceeding 1 kilogauss (see \citealp{Tsuneta08b, Ito10} for details). Moreover, \citet{Tsuneta08b} and \citet{Kaithakkal13} have reported that the kilogauss patches coincide in position with polar faculae at the photosphere. We expect the spatial intermittency of the magnetic field to persist even at the chromospheric level \textit{i.e.} with polar network bright points. In this paper, we investigate the association between the network bright points and magnetic patches using automated techniques. For this purpose, we have used SOT observations specifically devoted to observe polar regions of the Sun. In particular, we have used Ca~{\sc ii~H} passband images along with co-spatial and co-temporal level 2 spectro-polarimetric data observed by SOT. We have analyzed different datasets in polar regions of the Sun, details of which are given in Section~\ref{sec:data}. Results are discussed in Section~\ref{sec:result} and conclusions drawn from the study are summarized in Section~\ref{sec:conclusion}.


\begin{table}[htbp]	
	\caption{Summary of Ca~{\sc ii~H} observations from SOT.}
	\label{tbl1}
	\footnotesize
	\begin{tabular}{ccccc}
		\hline
		\bf{Data-Set} & \bf{Observation Time} &  \bf{FOV} &  \bf{Pointing (Xc,Yc)} & \bf{Target}\\
		\hline

		a & 2007-11-08 & $60^{\prime\prime}\times40^{\prime\prime}$ & $100^{\prime\prime},935^{\prime\prime}$  & North Pole \\
		& 13:04 to 13:58 UT & & & \\ 
		\hline
		
		b & 2008-12-04 &  $40^{\prime\prime}\times55^{\prime\prime}$  & $45^{\prime\prime},-933^{\prime\prime}$ & South Pole\\
		& 18:16 to 18:59 UT & & & \\
		\hline
		
		c & 2008-12-05 & $95^{\prime\prime}\times45^{\prime\prime}$  & $23^{\prime\prime},938^{\prime\prime}$ & North Pole\\
		& 10:01 to 10:59 UT & & & \\
		\hline
		
		d & 2009-04-01  &  $75^{\prime\prime}\times70^{\prime\prime}$ & $-138^{\prime\prime},-885^{\prime\prime}$ & South Pole\\
		& 12:00 to 12:59 UT & & & \\
		\hline
		
		e & 2010-05-14 & $160^{\prime\prime}\times40^{\prime\prime}$ & $20^{\prime\prime},-900^{\prime\prime}$  & South Pole \\
		& 12:02 to 12:57 UT & & & \\ 
		\hline

		f & 2012-03-18 & $150^{\prime\prime}\times50^{\prime\prime}$  & $5^{\prime\prime},-905^{\prime\prime}$ & South Pole\\
		& 13:00 to 13:59 UT & & & \\
		
		\hline
	\end{tabular}
\end{table}

\section{Observations and Data Analysis}\label{sec:data}
\subsection{Details of Observations}\label{sub:data1}

%


We have analyzed six polar datasets in this study (see Table~\ref{tbl1}). To study low chromospheric features, the Ca~{\sc ii~H} bandpass of \textit{Broad-Band Filter Imager} (BFI) of SOT, centered at  396.85 nm  with a band-width of 0.3 nm  is used. The SOT Ca~{\sc ii~H} data is calibrated by using the IDL routine of fg\_prep.pro available through SOT library of \textit{solarsoft}. For magnetic field estimation, we used co-spatial and co-temporal level 2 data obtained by  spectro-polarimeter(SP) of SOT. The SP creates high-precision Stokes polarimetric line profiles of the Fe I 6301.5 nm and 630.25 nm spectral lines. The primary product (Level~1) of the SP is Stokes IQUV spectra suitable for the derivation of vector magnetogram maps of the solar photosphere. The Level~2 data of the SP (provided at \url{https://csac.hao.ucar.edu/sp_data.php} and \url{http://sot.lmsal.com/data/sot/level2d/}) are outputs from spectral line inversions using the HAO "MERLIN" inversion code developed under the Community Spectro-polarimetric Analysis Center (CSAC) initiative. It performs a least-squares fitting of the Stokes profiles using the Milne-Eddington atmospheric approximation that allows for a linear variation of the source function along the line-of-sight, but holds the magnetic field vector, line strength, Doppler shift, line broadening, magnetic fill fraction (or scattered light fraction) constant along the line-of-sight. For every Ca~{\sc ii~H} data-set mentioned in Table~\ref{tbl1}, there exists a corresponding SP Level~2 data containing a complete set of inversion parameters. From the Level~2 data of SP, we mostly use map of magnetic field strength as our interest mainly lies in studying the relationship between the chromospheric Ca~{\sc ii~H} intensity and the strength of magnetic field present at the corresponding locations in the photosphere. Considering that the typical lifetime of supergranular cells is approximately a day and we are interested in the long-term enhanced heating of chromosphere, we performed all the analysis over time-averaged Ca~{\sc ii~H} images. Hence, we obtained a single average Ca~{\sc ii~H} intensity map for each dataset mentioned in Table 1. The pixel sampling of the Ca~{\sc ii~H} images is $0.11^{\prime\prime}$ whereas that of the maps magnetic field strength is $0.32^{\prime\prime}$ hence the pixel resolution of the Ca~{\sc ii~H} images was degraded to $0.32^{\prime\prime}$ (spatial resolution of $\sim0.65^{\prime\prime}$) in order to maintain the consistency between Ca~{\sc ii~H} intensity and magnetic filed strength maps throughout the study.

\begin{figure}[htbp]
	
	\centering 
	
	\includegraphics[width=0.9\textwidth]{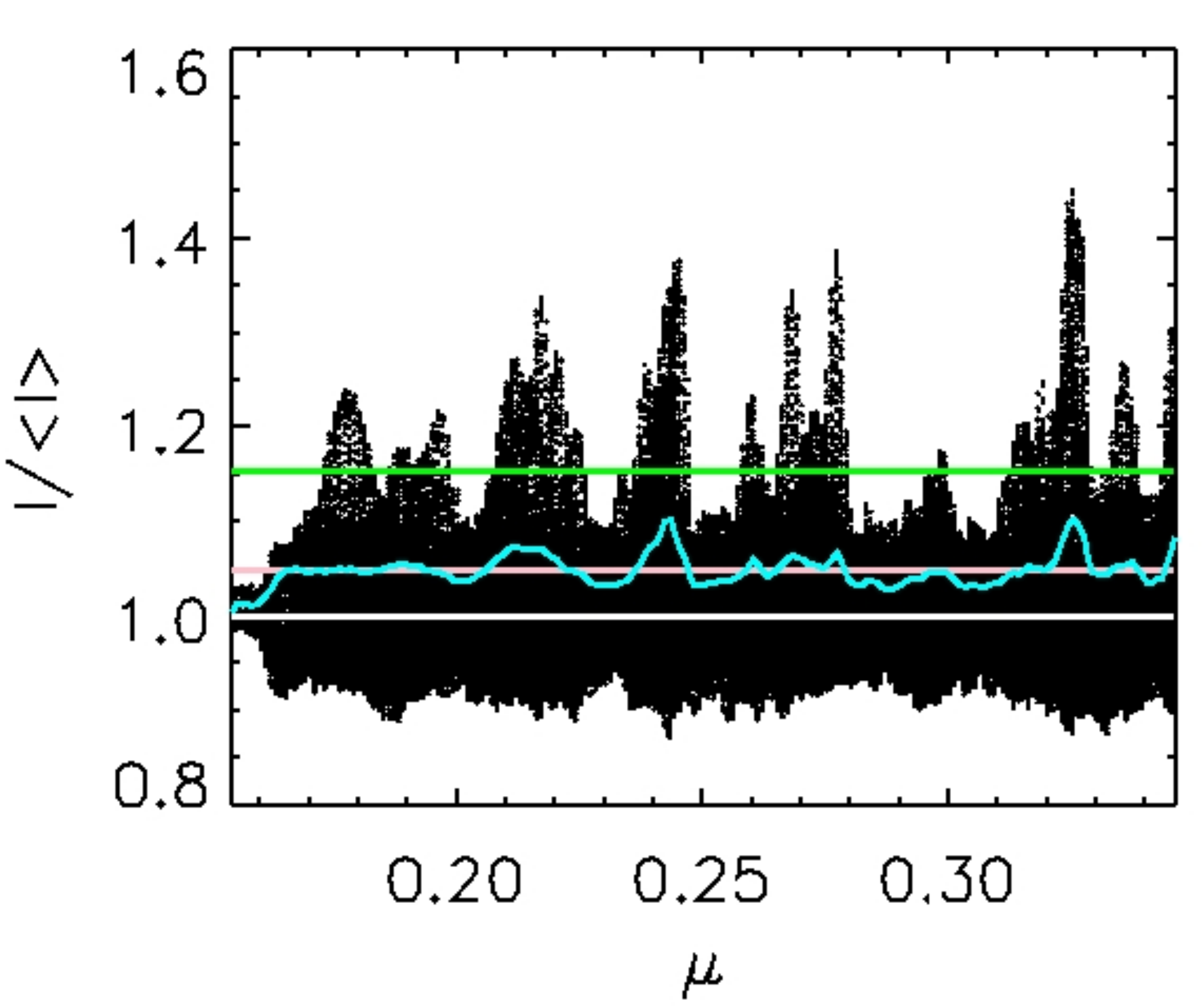}
	
	\caption{Normalised intensity ($ I / < I >$ ) \textit{vs} $\mu$ for all pixels of average Ca~{\sc ii~H} image of data-set (c) in Table 1. In this plot network bright points appear as intensity spikes. \textit{Blue curve} indicates the threshold for identifying network regions, \textit{pink line} marks the average of \textit{blue curve} and \textit{green line} indicates the threshold for selection of bright point pixels.}
	
	\label{fig2}
	
\end{figure}

\subsection{Identification of Bright-Point Regions}\label{sub:data2}

Pixels with the intensity greater than a given threshold (as discussed below) in Ca~{\sc ii~H} intensity map are classified as belonging to network bright points. The procedure applied to every average Ca~{\sc ii~H} image for selecting  such pixels is as follows (refer to Figure~\ref{fig1} and~\ref{fig2}): for every $\mu$, the mean and standard deviation($\sigma$) of the intensity over the corresponding pixels was calculated. The parameter $\mu$ is defined as the cosine of the angle between the surface normal and line of sight to the observer. It varies between 1 at disk center to 0 at the limb \citep{Thompson06}. For every $\mu$, the pixels with intensity greater than 1.1$\sigma$ of the respective mean were identified as network regions. Figure~\ref{fig2} shows the normalized intensity as a function of $\mu$ for the average Ca~{\sc ii~H} image of the data-set (c). The spikes in the plot are attributed to network bright points, while the smaller fluctuations about 1.0 are ascribed to network structure. The blue curve in Figure~\ref{fig2} indicates the threshold for network region selection. Figure~\ref{fig1} shows the average Ca~{\sc ii~H} normalized intensity image for the data-set (c) which marks the identified network regions enclosed in the blue contours.

From the selected network regions, the pixels with intensity more than 1.1 of average intensity of all the network regions selected in the image were marked as the locations of network bright points. In Figure~\ref{fig2}  the green line shows such threshold for detection of bright point pixels. The pink line marks the value of the average intensity over all the network regions detected in the image. The green line (threshold for the bright points) marks the value equals to 1.1 of the pink line. The detected bright point pixels were then clubbed into isolated bright point regions (using the IDL function \textit{label\_region}) with an area threshold of 50 pixels, \textit{i.e.} the bright point region should have the total number of pixels to be more than 50 in order to be selected. The spatial resolution of the images is $\sim0.65^{\prime\prime}$ and thus the area threshold for significant detection of the bright points should be more than $0.65^{\prime\prime}\times0.65^{\prime\prime}$. Henceforth, we fixed the area-threshold of 50 pixels which roughly corresponds to $1^{\prime\prime}\times1^{\prime\prime}$ area. Such detected regions were considered to represent the network bright point regions which are indicated by green contours in the Figure~\ref{fig1}. Every such identified network bright point region was then aligned with the corresponding region in the respective magnetogram. Figure~\ref{fig3} showcases various examples of co-aligned network bright points and magnetic patches where the network (blue) contours and bright point (green) contours are plotted over normalized Ca~{\sc ii~H} intensity and magnetic strength maps. Note that our method of identification of calcium network bright points is automated and thus is free from the involvement of human subjectivity. The limitation of our method lies in its inability to identify the bright points very close to the limb as noise becomes dominant because of low data-counts present near the limb.



\section{Results and Discussions}\label{sec:result}
\begin{figure}[htbp]
	
	\centering
	
	\includegraphics[width=0.9\linewidth]{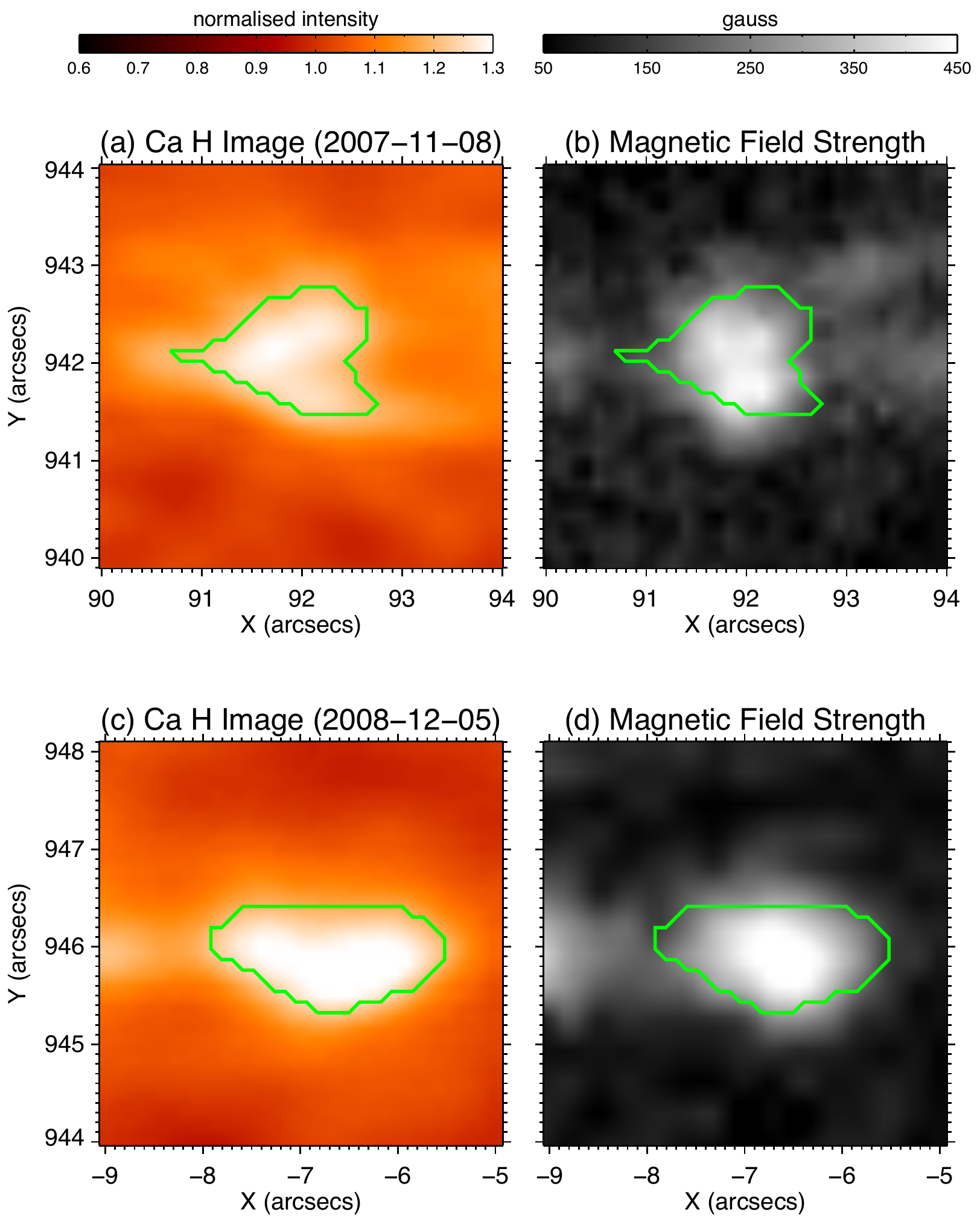}

	\caption{Representative examples showing association between network bright points and magnetic patches: (a) and (b)~Normalised SOT Ca~{\sc ii~H} image of a sub-region of data-set (a) and corresponding magnetic field map; (c) and (d) same as (a) and (b) but for data-set (c). The \textit{green contours} enclose the network bright-points detected from the Ca~{\sc ii~H} images as shown in the left panels. The right panels clearly show that the \textit{green contours}, which represent the network bright points, coincide with the magnetic patches of high field strengths.}
	
	\label{fig3}
	
\end{figure}

\begin{figure}[htbp]
	
	\centering
	
	\includegraphics[width=0.9\textwidth]{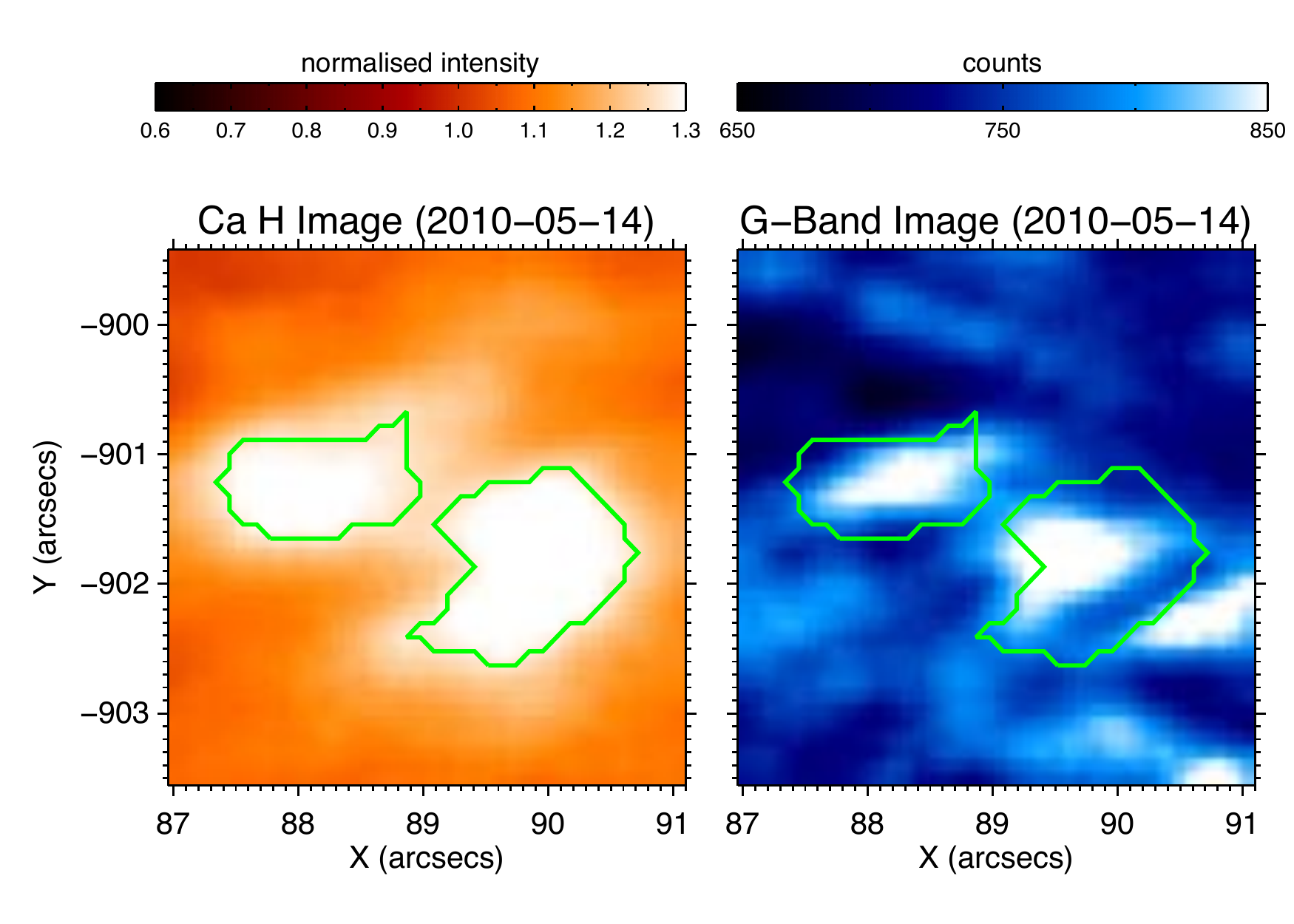}

	\caption{Ca~{\sc ii~H} and G-band image of a sub-region of data-set (e), showing relationship between the Ca~{\sc ii~H} bright-points with photospheric faculae. The \textit{green contours} are representing the network bright points which happen to generally lie over the groups of g-band bright points.}
	
	\label{fig4}
	
\end{figure}

We find that the network bright points, as seen in Ca~{\sc ii~H} images are almost always associated with the magnetic patches as seen in the photosphere (as illustrated in Figure~\ref{fig3}). Their sizes are $\sim1^{\prime\prime}-5^{\prime\prime}$. The lower limit of $1^{\prime\prime}$ of size of the bright-points is dictated by the minimum area-threshold applied while detecting the bright-points (mentioned in Section~\ref{sub:data2}). As shown in a particular example in Figure~\ref{fig4}, the Ca~{\sc ii~H} bright-points (as defined here) seem to be co-spatial with the groups of G-band bright-points or faculae. Figure~\ref{fig1}~and~\ref{fig3} clearly show that the percentage area covered by the network bright points (green contours) is considerably small. Table~\ref{tbl2} shows the percentage area of bright points in each data-set studied, along with average magnetic field strength over the whole field of view and  only within bright points. Though the percentage area occupied is low, the magnetic field strength concentrated under the network bright points is generally high as compared to the values of the average field strength over the full FOV of SOT observations. This implies that the overall behaviour of the polar magnetic field should be governed by the magnetic fields underneath these locations on an average.

It is important to note that we do not consider all the pixels of the FOV while performing the averaging over the entire FOV. The pixels where the inversion code could not converge (reflected as \textit{NAN} values) were completely discarded while taking the average. Such locations generally occur in the internetwork regions where the signal is poor to obtain any reliable fits to the Stokes profiles. The internetwork locations could possibly be having lower magnetic field strengths~\citep{Lites08} which cannot be estimated due to observational and technical limitations in case of polar region observations. The exclusion of such locations can effect the value of the average field strength calculated over the full FOV. On the other hand, this do not effect the estimation of magnetic field strength in the bright-points as these are significantly brighter regions with considerable high signal.

Figure~\ref{fig4} shows that the value of the magnetic field strength can reach up to 600 gauss within the bright point regions. The maximum field strength within the bright point, which happen to be at the centre of the bright point/magnetic patch, reaches up to 1 kilogauss in many cases, as shown in Table~\ref{tbl2}. \citet{Tsuneta08b} and \citet{Kaithakkal13} have also reported the values of the order of a few kilogauss in concentrated magnetic patches. More recently, \citet{Pastor18} have also observed the presence of strong magnetic field elements of strength 600-800 gauss at the solar poles. In the past era, the low-resolution observations were unable to detect these small-scale locations of strong magnetic field strengths which is now possible with enhanced resolution data.

\begin{figure}[htbp]

	\centering

	\includegraphics[width=0.9\textwidth]{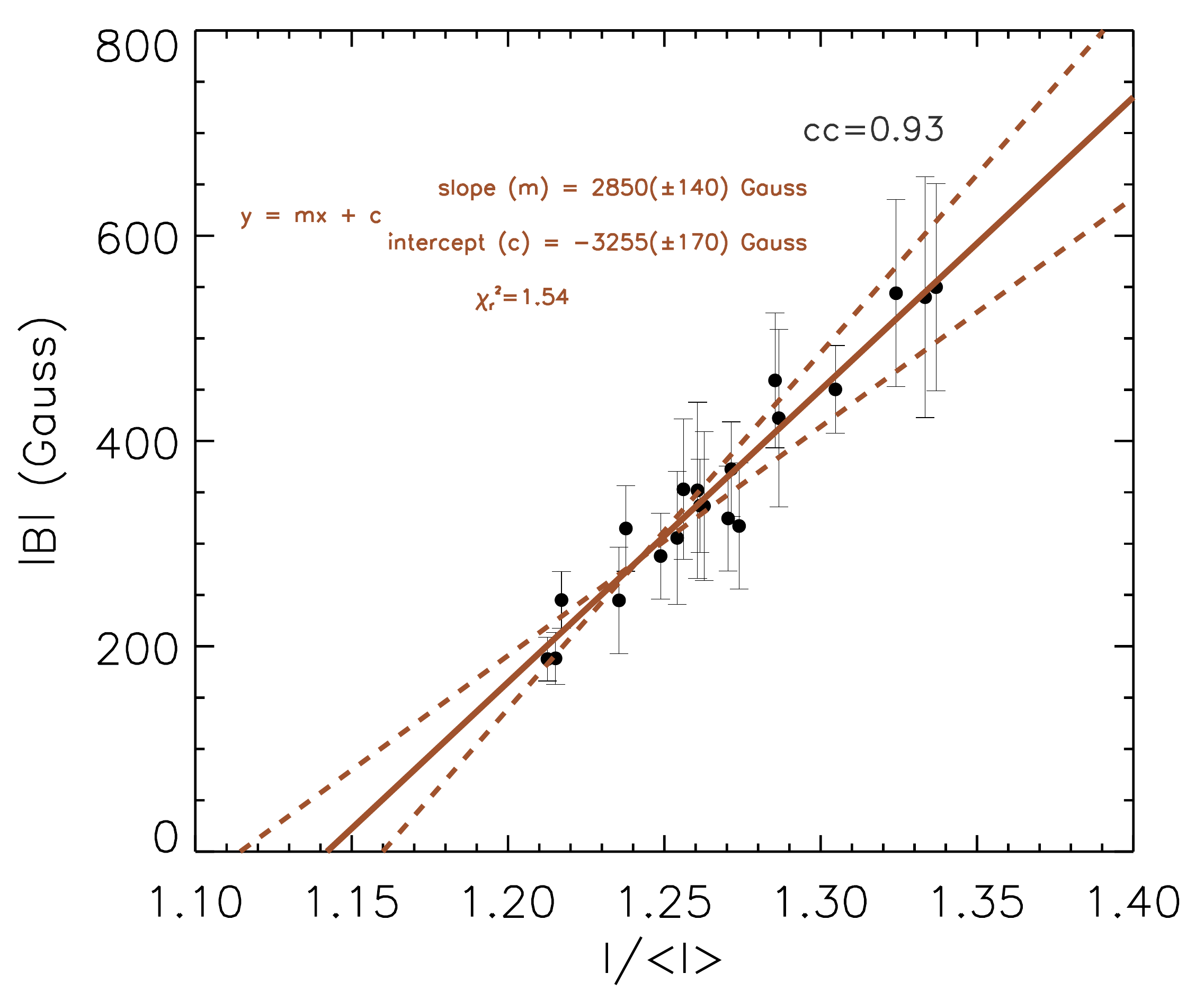}

	\caption{Scatter plot between normalised Ca~{\sc ii~H} intensity ({\scriptsize$I~/<I>$}) and magnetic field strength ({\scriptsize $|B|$}) for all the 20 bright points  detected in the six data-sets. Every point in the plot represents the respective values averaged within each bright point region. cc indicates the correlation coefficient, m indicates the slope and c indicates the intercept.}
	
	\label{fig5}
	
\end{figure}

Along with considerable spatial association observed between network bright points and magnetic patches, we find that a good correlation exists between bright point intensity and magnetic field strength. Figure~\ref{fig5} shows the scatter plot of the normalised Ca~{\sc ii~H} intenisty ({\scriptsize $I~/<I>$}) and magnetic field strength ({\scriptsize $|B|$}) for all the bright points detected in the 6 data-sets. Note the high value of correlation coefficient (cc) of 0.93. We have attempted to calibrate the normalized Ca~{\sc ii~H} intensity with the corresponding magnetic field strength of the bright points. In Figure~\ref{fig5}, the straight solid line shows the best fit (based on the chi-square minimization) and corresponding slope and intercept is indicated by \textit{m} and \textit{c} along with the respective fitting errors in the estimates. For a network bright point with the value of {\scriptsize $I~/<I>$} that is greater than 1.14, the value of {\scriptsize $|B|$} can be calculated by using this value of the slope (m=2850 gauss) and the intercept (c=--3255 gauss). For {\scriptsize $I~/<I>$}  less than 1.14, the deduced relation will yield a negative value of {\scriptsize $|B|$} which is unphysical. Considering the errors in the slope and intercept, the percentage error in the deduced magnetic field strength from this linear relation turns out be 10\%.

It is worthwhile to mention that the magneto-hydrodynamic (MHD) models by~\citet{Schrijver89a,Solanki91} and~\citet{Schrijver93} have predicted the presence of power-law dependence of calcium intensity over the magnetic field strength as shown in Equation~\ref{eq1} below:
\begin{equation}
I = I_{0}+M|B|^{a}
\label{eq1}
\end{equation}

\noindent
where $I$ denotes the observed calcium intensity, $I_{0}$ represents the \textit{zero or basal}  intensity, $|B|$ denotes magnetic field strength, $M$ is a constant multiplier and $a$ to be the power-law index. For instance,  the flux tube model of~\citet{Schrijver93} have proposed the value of the power-law index ($a$) to be $\sim0.4$. Whereas, different observations~\citep{Shrijver89b,Harvey99,Rezaei07,Louki09} have obtained different values of \textit{a} to be ranging from 0.2 to 0.7. On the other hand,~\citet{Skumanich75} and ~\citet{Nindos98} have observed the linear relation (\textit{i.e.} $a=1.0$), which yields similar result as  obtained in the present study.

\begin{table}[htbp]
	
	\caption{Properties of network bright points.}
	\label{tbl2}
	\small
	\begin{tabular}{ccccc}
		
		\hline
		
		\bf{Data-} & \bf{\% area}  & \bf{avg. $\bf|B|$}  & \bf{avg. $\bf|B|$} & \bf{max. $\bf|B|$}\\
		
		\bf{set} & \bf{covered by}  &\bf{over}   & \bf{over all} & \bf{within the} \\
		
		&  \bf{bright points} & \bf{full FOV} &  \bf{bright points} &  \bf{bright points}\\
		
		\hline

		a & 0.2 & 93 gauss & 320 gauss & 695 gauss  \\
		
		\hline
		
		b & 0.2 &  92 gauss  & 267 gauss &  634 gauss\\
		
		\hline
		
		c& 0.5 & 89 gauss  & 292 gauss &  983 gauss\\
		
		\hline
		
		d & 0.6  & 94 gauss  & 392 gauss & 934 gauss\\
		
		\hline
		
		e & 0.8 & 107 gauss & 500 gauss & 1210 gauss\\
		
		\hline

		f & 0.5 & 122 gauss  & 322 gauss & 686 gauss\\

		\hline
		
	\end{tabular}
\end{table}



The linear relation between Ca~{\sc ii~H} intensity and magnetic field strength indicates that the network bright-points can be considered as individual flux-tubes \citep{Shrijver89b}, well isolated from each other. The atmospheric heating over such flux-tubes is independent of the ambient magnetic topology and hence results in the linear relation between Ca~{\sc ii~H} (or K) intensity and field strength. In addition, as pointed out by~\citet{Skumanich75}, the non-linear effects appear with the inclusion of weak field ($<50$\,gauss) and very strong field ($>1$\,kilogauss) locations. The linear trend obtained here could be an effect of selection of the features of interest. Hence, it is important to note that the linear relation obtained here is valid generally for the network bright points \textit{i.e.} for the compact brighter locations ($\sim1^{\prime\prime}-5^{\prime\prime}$) inside the bright network regions with the values of normalised calcium intensity to be more than 1.20 and thus magnetic field strengths to be greater than 200 gauss, up to $\sim1$\,kilogauss (as depicted by Figure~\ref{fig5}). Beyond this regime, which could map different class of features and structures, this linear relation may or may not hold good,~\textit{i.e.} the power-law index (\textit{a}) in Equation (1) could be different than 1.0 with completely different values of other constants. Moreover, for the active region plages, which posses similar magnetic field strengths, intensity values and sizes but are embedded in a completely different magnetic environment as compared to network bright points, the relation between calcium intensity and magnetic field strength could be different (see~\citealp{Harvey99} for details).

\section{Conclusions}\label{sec:conclusion}

 Recently, \citet{Tsuneta08b} and \citet{Kaithakkal13} have shown that the magnetic patches of strengths $\sim$1~kilogauss coincide in position with polar faculae. In the present work,  we have studied six polar region observations from SOT and established that such an association between enhanced brightness and magnteic field strength persists up to the chomosphere in the polar regions of the Sun. We find that the calcium network the bright points mostly exists at the locations of concentrated magnetic field locations. We have observed a considerable spatial association between the network bright points and magnetic patches in all the data-sets. Moreover, a good correlation exists between normalized Ca~{\sc ii~H} intensity and photospheric magnetic field strength of the network points and linear relation is present between them. Such linear trend indicates that the network bright-points can be regraded as isolated flux-tubes anchored in the photosphere, and the atmospheric heating above these locations is insensitive to the neighbouring magnetic environment. Though the percentage area covered by the network bright points is considerably small, they posses high magnetic field values and thus contribute to the global polar field majorly. These chromospheric Ca~{\sc ii~H} bright points seems to be co-spatial with groups of G-band bright points in th photosphere. This clearly indicates that these different features are directly coupled with each other, though present in different layers of the solar atmosphere, and happen to be manifestations of the magnetic field concentrations present in the lower photosphere.

\section*{Acknowledgments}
\textit{Hinode} is a Japanese mission developed and launched by ISAS/JAXA, with NAOJ as domestic partner and NASA and STFC (UK) as international partners. It is operated by these agencies in cooperation with ESA and NSC (Norway). N.\,N. acknowledges Human Resource Development Group (HRDG) of Council of Scientific and Industrial Research (CSIR), India for awarding Senior Research Fellowship (SRF). V.\,P. is supported by European Research Council (ERC) under the European Union's Horizon 2020 research and innovation programme. Authors would like to thank anonymous referee for her/his detailed comments which led to the improvement of the manuscript.


\end{article} 

\end{document}